\documentclass[english,11pt]{elsarticle}
\usepackage{etex}
\usepackage[english]{babel}
\usepackage{geometry}
\usepackage{graphics}
\usepackage{amssymb}
\usepackage{amsmath}
\usepackage{url}
\usepackage{epsfig}
\usepackage{latexsym}
\usepackage{t1enc}
\usepackage{amsmath}
\usepackage{amsthm}

\usepackage{times}
\usepackage{pifont}
\usepackage{fancybox}
\usepackage{algpseudocode}
\usepackage{fancyhdr}
\usepackage{graphics}
\usepackage{lscape}
\usepackage{booktabs}
\usepackage{algorithm}
\usepackage{algpseudocode}
\usepackage{tikz,pgfplots}
\usepackage{amsthm}
\usepackage{pgf}
\usepackage{tikz,tkz-graph}
\usetikzlibrary{arrows,plotmarks}
\usetikzlibrary{arrows,automata}
\usepackage{stmaryrd}
\usepackage{longtable}
\usepackage{subfig}

\numberwithin{figure}{section}
\newtheorem{theorem}{Theorem}[section]

\newtheorem{example}[theorem]{Example}
\allowdisplaybreaks

\begin{document}
\begin{sloppypar}

\begin{frontmatter}

%
\title{A Rewriting System for the Assessment of XACML Policies Relationship}

\author{M. Mejri$^1$, H. Yahyaoui$^2$, A. Mourad$^3$, and M. Chehab$^3$\\
\small{$^1$Computer Science Department, Laval University, Canada}\\
\small{{\em mejri@ift.ulaval.ca}}\\
\small{$^2$Computer Science Department, Kuwait University, Kuwait}\\
\small{{\em hamdi@cs.ku.edu.kw}}\\
\small{$^3$Department of Computer Science and Mathematics, Lebanese American University, Lebanon}\\
\small{{\em azzam.mourad@lau.edu.lb},{\em lemirmohammad.shehab@lau.edu}}}

\begin{abstract}
We propose in this paper a new approach to assess the relationship between XACML policies. Our approach spans over three steps. In the first one, the XACML policies are mapped to terms of a boolean ring while taking into account XACML policy and rule combining algorithms.  In the second step, the relationship problem between XACML policies is transformed into a validity problem in a boolean ring. In the third step, the validity problem is resolved using a dedicated rewriting system. The convergence of the rewriting system is proved in this paper. Moreover, the approach is implemented and its performance is evaluated. The results show that our approach enjoys better performance and memory cost than the best so far published SMT based approach.
\end{abstract}
\end{frontmatter}

\section{Introduction}

Assessing the relationship between policies is inevitable and crucial in different situations. For instance, it could be illegal for a company or a governmental agency to store some kind of data (health or tax records, etc.) with cloud service providers that do not comply with a given privacy policy. A cloud server that can search for commercially valuable data in user storage space or that gives itself the right to change the user security policy may threaten the user data privacy. The discrepancy between user and provider security policies can create security challenges \cite{Gellman2009PrivacyIT}. Accordingly, several security policy languages were devised for the sake of policies specification and analysis. Standard security policy languages are designed to create a common background for providers to define their security constraints. A paramount benefit from such standardization is the interoperability between different systems. Generally, the proposed security languages come with an informal syntax and semantics, which makes the learning curve of such languages high and the policies specification error-prone. One of these languages is eXtensible Access Control Markup Language (XACML) \cite{ref9}. It is a de facto standard specification language used to define and enforce access control policies for different kinds of systems including cloud computing systems, internet of things and blockchain access control \cite{AAMAWW2018,DBLP:conf/dais/MaesaMR17,Xu:2018:ACT:3205977.3205979}. Next Generation Access Control (NGAC) \cite{NGACFA2013,GOADS2016, FCHK2016-1} is also an interesting attribute-based access control language. A good comparison between XACML and NGAC has been done by Ferraiolo et al. in \cite{FCHK2016}.

%

Policies analysis requires establishing a formal framework that allows reasoning about policies. We focus in this paper
on policies relationship assessment using a formal approach. More particularly, our objective is to devise a framework that allows users to assess the relationship between two security policies such as equivalence, restriction, inclusion, and divergence. Our work aims to achieve that objective for the standard de facto XACML language. Thus, we formalize in this paper XACML policies with their combining algorithms. More
precisely, we transform XACML policies to terms of a boolean ring. Then, we use a rewriting system to assess the policies relationship.

Compared to similar proposed approaches such as \cite{DBLP:conf/post/TurkmenHRZ15,DBLP:journals/compsec/TurkmenHRZ17} and \cite{DBLP:journals/corr/MargheriPT15}, our work comes with a transformation of XACML policies to compact boolean terms of a boolean ring and hence enjoys better readability and learning curve than these approaches as will be discussed later in the related work section. Furthermore, using a rewriting system to prove the relationship between XACML policies gives us an extensible framework that can address more problems than what Satisfiability Modulo Theories (SMT) and Binary Decision Diagrams (BDD) can do. Also, as shown in \cite{Dershowitz04booleanring}, an SMT based on a boolean ring and rewriting can be efficient in many cases and it even gives polynomial-time results for restricted classes of boolean ring formulae. The results show that our approach enjoys better performance and memory cost than the best so far published SMT based approach \cite{DBLP:journals/compsec/TurkmenHRZ17}.

The contributions of this work are as follows:

\begin{itemize}
\item The transformation of XACML policies into terms of a boolean ring.

\item The proposal of a rewriting system for security policies that can be used to establish the
relationship between their semantic terms.

\item The evaluation of the proposed approach.
\end{itemize}

The rest of the paper is organized as follows. An overview about XACML is provided in Section \ref{xacmloverview}. The background related to
boolean rings, boolean algebra and rewriting systems is presented in Section \ref{BooleanAlgebra}. Section \ref{approach} is dedicated to the presentation of the proposed approach and the termination of the rewriting system. In section \ref{exp}, we provide an experimental evaluation of the proposed approach. In Section \ref{relwork}, we discuss the related work regarding policies relationship. Finally, some concluding remarks and future work are provided in Section \ref{conc}.

\section{Overview of XACML}
\label{xacmloverview}

\begin{figure}[!h]
    \centering
    \includegraphics[scale=0.5]{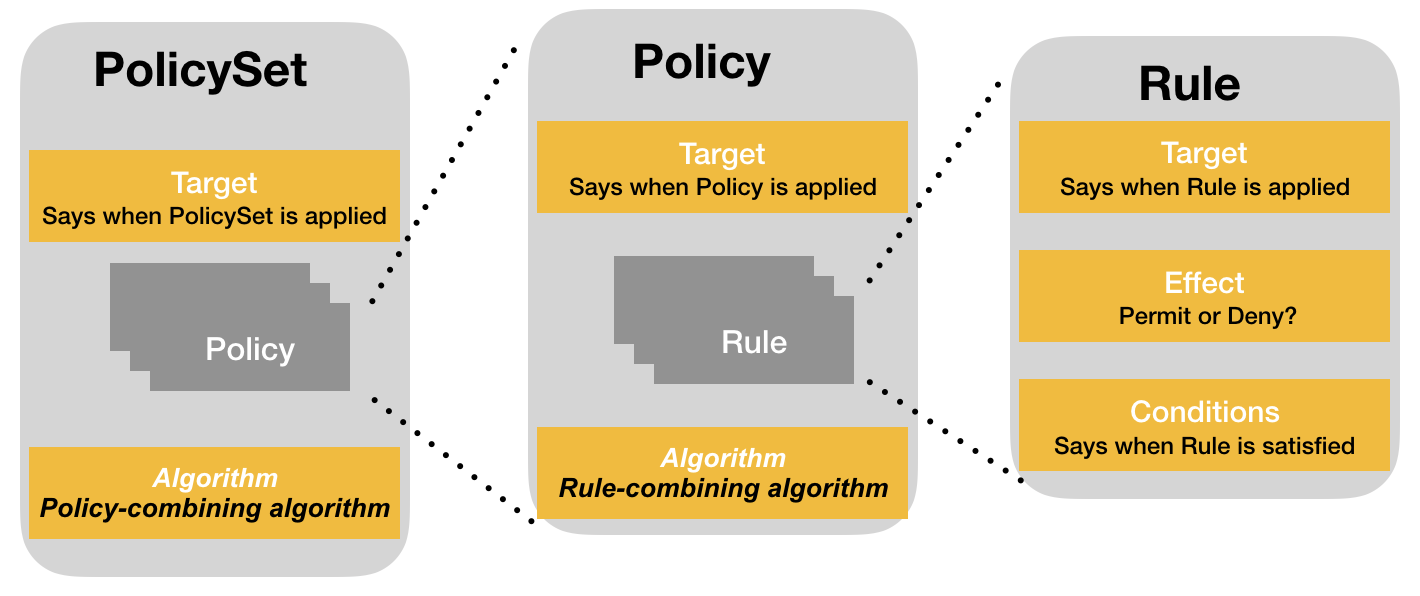}
    \caption{XACML structure}
    \label{Xstru}
\end{figure}

XACML (eXtensible Access Control Markup Language) \cite{ref9} is a set of XML schemas that define the specification of a language for access control policies. As shown in Figure~\ref{Xstru}, an XACML policy is composed of a set of rules. Rules are also composed of target conditions and effects or permissions. Given a request, the goal of each rule is to make a decision (permit or deny).

Because a policy may contain multiple rules with different decisions,  we need to clarify how to build the decision of a policy from the decision of its rules.  To this end,  XACML comes with ``Rules Combining Algorithms". It is also possible to aggregate policies to form a \textit{PolicySet}. Like policies, a \textit{Policyset} has also a target that limits the scope of its applicability and an algorithm that defines its global decision from the local decisions returned by its policies. The target of an access request is first compared to the target of a \textit{Policyset}. If they are different, this \textit{Policyset} is not applicable. Otherwise, the target of the request is compared to the targets of policies inside the \textit{Policyset}. A policy is qualified as not applicable when its target is different from the target of the request. When a policy and a request agree on the target, the request is analyzed by the rules inside the policy. A rule is applicable if its targets include the target of the request and its condition is evaluated to true.

The XACML standard comes with four types of combining algorithms. Hereafter, we describe how a \textit{PolicySet} combines the results of its policies. The same algorithms can be used to combine the decisions of rules to build the decision of their policies.

\begin{itemize}
    \item \textbf{Permit-overrides}: A PolicySet accepts if at least one of its policies accepts and denies if no one of its policies accepts and at least one denies.  Otherwise, the PolicySet is not applicable.
    \item \textbf{Deny-overrides}: It returns ``deny" if at least one policy denies and it returns ``accept"  if no policy denies and at least one policy accepts. Otherwise, the policy is not applicable.
    \item \textbf{First Applicable}: It returns ``accept" if there is at least one policy that accepts and this policy is not preceded by a denying one and vice versa.
    \item \textbf{Only-one-applicable}: If more than one policy is applicable, then the result is indeterminate. Otherwise, the result of the unique applicable policy will be considered.
\end{itemize}

The BNF grammar in Table \ref{TabXACML} gives more details about the
full XACML 3.0 language except the following combining algorithms: ordered-permit-overrides and
ordered deny-override. Some of the operators such as string comparison operators are omitted in
this grammar but can be handled in the same way as the integer comparison operators, which are
taken into account in this work.

\begin{table*}[!h]
    \begin{center}

        \scalebox{0.43}{
            $\begin{array}{||lll||}
            \hline
            \hline
            PDPpolicies &::=& PolicySet \mid Policy\\
            &&\\
            PolicySet&::=& \boldsymbol{<}\textbf{POLICYSET}\;\; Pheader\; \boldsymbol{>}  \;\;[Description] \;\;Targets\;\; Policies  \;\;  [Obligation] \;\; [Advice] \;\; \boldsymbol{<\!/}\textbf{POLICYSET } \boldsymbol{>}  \\
            Policy& ::=&  \boldsymbol{<}\textbf{POLICY }\;\;  Rheader\;\boldsymbol{>} \;\;[Description] \;\;Targets\;\; Rules \;\;  [Obligation] \;\; [Advice]  \;\; \boldsymbol{<\!/}\textbf{POLICY} \;\;\boldsymbol{>}  \\
            Policies &::=& Policy\mid Policy \;\; Policies\\
            Rules &::=& \boldsymbol{<}\textbf{RULE} \;\;  Rheader\;\boldsymbol{>} \;\; [Description]\;\; [Targets] \;\;[Condition] \;\;  [Obligation] \;\; [Advice] \boldsymbol{<\!/}\textbf{RULE}\;\;\boldsymbol{>}\\
            &&\\
            PSheader & ::=&\textbf{PolicySetId} \;\;\boldsymbol{=}\;\; string\;\; \textbf{Version} \;\;\boldsymbol{=}\;\; number\;\; \textbf{ PolicyCombiningAlgId}\;\;\boldsymbol{=}\;\;Palg\\
            Pheader & ::=&\textbf{PolicyeId}\;\; \boldsymbol{=} \;\; string\;\; \textbf{Version} \;\;\boldsymbol{=}\;\; number\;\; \textbf{ RuleCombiningAlgId}\;\; \boldsymbol{=}\;\;Ralg\\
            Rheader & ::=&\textbf{RuleId}\;\; \boldsymbol{=} \;\; string\;\; \textbf{Effect}\;\; \boldsymbol{=}\;\;REffect\\
            &&\\
            Palg & ::= & \textbf{only-one-applicable} \\
            & & \mid  \textit{Ralg} \\
            &&\\
            Ralg & ::= & \textbf{deny-overrides} \\
            & & \mid \textbf{permit-overrides}\\
            & & \mid \textbf{first-applicable}\\
            & & \mid \textbf{ordered-permit-overrides}\\
            &&\\
            REffect &::=& \textbf{Permit} \mid \textbf{Deny} \\
            &&\\
            Targets &::=& \boldsymbol{<}\textbf{TARGET}\boldsymbol{>}\;\; [MatchAny] \boldsymbol{</}\textbf{TARGET}\boldsymbol{>}\\
            &&\\
            MatchAny &::=& \boldsymbol{<}\textbf{AnyOf }\boldsymbol{>} matchAll \boldsymbol{</}\textbf{AnyOf }\boldsymbol{>} \\
            & & \mid \boldsymbol{<}\textbf{AnyOf }\boldsymbol{>} matchAll \boldsymbol{</}\textbf{AnyOf }\boldsymbol{>} MatchAny\\
            &&\\
            MatchAll &::=& \boldsymbol{<}\textbf{AllOf }\boldsymbol{>} Matches \boldsymbol{</}\textbf{AnyOf }\boldsymbol{>} \\
            & & \mid \boldsymbol{<}\textbf{AnyOf }\boldsymbol{>} Matches \boldsymbol{</}\textbf{AllOf }\boldsymbol{>} MatchAll\\
            &&\\
            Matches &::=& Match \mid Match\;\; Matches\\
            &&\\
            Match &::=& \boldsymbol{<}\textbf{Match } \textbf{MatchID = } MatchId\boldsymbol{>}\;\;\\
            && \hspace{0.4cm} \boldsymbol{<}\textbf{AttrValue}\boldsymbol{>} value \boldsymbol{</}\textbf{AttrValue}\boldsymbol{>}\\
            && \hspace{0.4cm} \boldsymbol{<}\textbf{AttributeDesignator }\;\;ADHeader \boldsymbol{/>}\\
            && \hspace{0.2cm} \boldsymbol{</}\textbf{Match } \boldsymbol{>}\\
            &&\\
          MatchId & ::= & \textbf{string-equal} \\
            & & \mid \textbf{integer-equal}\\
            & & \mid \textbf{string-regexp-match} \\
            & & \mid \textbf{integer-greater-than}\\
            & & \mid \textit{\ldots}\\
            &&\\
            ADHeader &::=&  \textbf{Category= } Subject \;\; \textbf{AttributeId = } AttSubject \;\;\textbf{DataType= } type\;\; \textbf{MustBePresent= } \textit{boolean} \\
            & & \mid \textbf{Category= resource} \;\; \textbf{AttributeId = } AttResource \;\;\textbf{DataType= } type\;\; \textbf{MustBePresent= } \textit{boolean} \\
            & & \mid \textbf{Category= action} \;\; \textbf{AttributeId = } AttAction \;\;\textbf{DataType= } type\;\; \textbf{MustBePresent= } \textit{boolean} \\
            & & \mid  \textbf{Category= environment} \;\; \textbf{AttributeId = } AttEnv \;\;\textbf{DataType= } type\;\; \textbf{MustBePresent= } \textit{boolean} \\
            &&\\
            Subject & ::= & \textbf{access-subject} \\
            & & \mid \textbf{recipient-subject} \\
            & & \mid  \textbf{intermediary-subject}\\
            && \ldots\\
            &&\\
            AttSubject & ::= & \textbf{subject-id} \\
            & & \mid \textbf{subject-id-qualifier} \\
            & & \mid  \textbf{key-info}\\
            & & \mid  \textbf{authentication-time}\\
            && \ldots\\
            &&\\
            AttResource& ::= & \textbf{resource-id} \\
            & & \mid \textbf{target-namespace} \\
            &&\\
            AttAction& ::= & \textbf{action-id} \\
            & & \mid \textbf{implied-action} \\
            & & \mid \textbf{action-namespace} \\
            &&\\
            AttEnv& ::= & \textbf{current-time} \\
            & & \mid \textbf{current-date} \\
            & & \mid \textbf{current-dateTime} \\
            &&\\
            type & ::= & \textbf{x500Name} \\
            & & \mid \textbf{rfc822Name} \\
            & & \mid \textbf{ipAddress} \\
            & & \mid  \textbf{dnsName}\\
            & & \mid  \textbf{xpathExpression}\\
            & & \mid  \textbf{string}\\
            & & \mid  \textbf{boolean}\\
            & & \mid  \textbf{double}\\
            & & \mid  \textbf{time}\\
            & & \mid  \textbf{date}\\
            & & \mid  \textbf{dateTime}\\
            & & \mid  \textbf{anyURI}\\
            & & \mid  \textbf{hexBinary}\\
            & & \mid  \textbf{base64Binary}\\
            &&\\
            Condition &::=& \boldsymbol{<}\textbf{ Condition}\boldsymbol{>} BooleanExpression  \boldsymbol{</}\textbf{ Condition}\boldsymbol{>}\\
            \hline
            \hline
        \end{array}$
    }
     \caption{{\scriptsize A BNF Grammar for a Subset of XACML-3.0}}
     \label{TabXACML}
    \end{center}
    \end{table*}

\section{Overview of Boolean Rings, Boolean Algebras and Rewriting Systems}
  \label{BooleanAlgebra}

A boolean ring ($B$, $\oplus$, $\ast$, 0, 1) is a commutative ring in which $+$ is nilpotent ($x\oplus x=0$) and $\ast$ is idempotent ($x\ast x= x$). There is an isomorphic relation between a boolean ring and a boolean algebra, i.e., we can build a boolean ring form a boolean algebra and vice versa.

 \begin{itemize}
 	\item From each boolean ring, we can construct a boolean algebra:\\
 	If ($B$, $\oplus$, $\ast$, 0, 1) is a boolean ring, then ($B$, $\vee$, $\wedge$, $\neg$, 0, 1)
 	is a boolean algebra such that $\vee$, $\wedge$, $\neg$ are defined as shown in Table \ref{tab:ringtoalg}.
 	\begin{table}[H]
 		\caption{From Rings to Algebras} \label{tab:ringtoalg}
 		\begin{center}
 			$\begin{array}{lll}
 			x \vee y&= & x \oplus y \oplus x \ast y  \\
 			x \wedge y &=& x \ast y\\
 			\neg x  &= & x \oplus 1 \\
 			\end{array}$
 		\end{center}
 	\end{table}
 	
 	\item Inversely, from a boolean algebra we can construct a boolean ring:\\
 	If ($B$, $\vee$, $\wedge$, $\neg$, 0, 1) is a boolean algebra, then ($B$, $\oplus$, $\ast$, 0, 1)
 	is a boolean ring such that $\oplus$ and $\ast$ are defined in Table \ref{tab:algtoring}.
 	\begin{table}[H]
 		\caption{From Algebras to Rings} \label{tab:algtoring}
 		\begin{center}
 			$\begin{array}{lll}
 			x \oplus y&= & (x \wedge \neg y) \vee (\neg x \wedge y)  \\
 			x \ast y &=& x \wedge y\\
 			\end{array}$
 		\end{center}
 		
 	\end{table}
 \end{itemize}

Unlike boolean algebra formalism, the boolean ring formalism defines a unique normal form allowing to transform any term to a unique form, up to associativity and commutativity of the two operators $\ast$ and $\oplus$, thanks to the nilpotence of $\oplus$ and the idempotence of $\ast$. This form is 1, 0, or a sum ($\oplus$) of distinct monomials (either $1$, a variable $x$ or a product of many variables, e.g., $x_1 \ast x_2\ast x_5$).  We can eliminate redundant variables in a monomial using the idempotence property and we can eliminate the redundant monomials using the nilpotence property until we reach a normal form (a form that cannot be further simplified).  Since any boolean formula can be transformed to a term of a boolean ring, then it can benefit from the simplification properties of boolean ring operators to be reduced to a unique normal form.  To make this simplification automatic, we usually use a rewriting system, i.e., a set of rules $l_i\rightarrow r_i$, $i=1,\ldots, n$,  that can be applied to transform any term $t$ such that $t=\sigma(l_i)$ to $\sigma(r_i)$, where $\sigma$ is a substitution.  If this simplification process always terminates for any starting term $t$ and always leads to a unique simplified form of $t$, the rewriting system is called canonical or convergent. A canonical Rewriting system for Boolean Algebra, denoted by RBA, is given in Table \ref{MainRule}, where ${\sf F}$ and ${\sf T}$ is an alternative representation of true and false.

  \begin{table}[!h]
  	\caption{Main Inference Rules\label{MainRule}}
  	\begin{center}
  		$\begin{array}{llll}
  		{\sf F}&\rightarrow &0& R_{0}\\
  		{\sf T}
  		&\rightarrow &1 & R_{1}\\
  		x \vee y
  		& \rightarrow &x \ast y \oplus x \oplus y &  R_{2}\\
  		x \wedge y
  		&\rightarrow& x \ast y & R_{3}\\
  		x \Longrightarrow y
  		&\rightarrow &x \ast y \oplus x \oplus 1 & R_{4}\\
  		x \approx y
  		&\rightarrow &
  		x \oplus y \oplus 1
  		& R_{5}\\
  		\neg x
  		&\rightarrow &x \oplus 1& R_{6}\\
  		x \oplus 0
  		&\rightarrow& x& R_{7}\\
  		x \oplus x
  		&\rightarrow &0& R_{8}\\
  		x \ast 1
  		&\rightarrow &x  & R_{9}\\
  		x \ast x
  		&\rightarrow &x & R_{10}\\
  		x \ast 0
  		&\rightarrow&0  & R_{11}\\
  		x \ast (y \oplus z)
  		&\rightarrow& x \ast y \oplus x \ast z &R_{12}\\
  		\end{array}$
  	\end{center}
  \end{table}

The RBA system can be used as a theorem prover to check if a boolean term is a tautology (it can be reduced to 1), a contradiction (it can be reduced to 0), or a satisfiable formula.
	
	$$
	\begin{array}{ll}
	&p\wedge (p\vee q)\\
	\rightarrow& (R_4)\\
	&p\ast  (p\vee q)\\
		\rightarrow& (R_3)\\
		&	p\ast  (p\ast q \oplus p \oplus q)\\
			\rightarrow& (R_{13})\\
		&	p\ast p\ast q \oplus p\ast p \oplus p\ast q\\
					\rightarrow& (R_{11})\\
		&	 p\ast q \oplus  p \oplus p\ast q\\
		
							\rightarrow& (R_{9})\\
		&	   p \oplus 0\\
									\rightarrow& (R_{8})\\
		&	   p \\
	\end{array}
	$$
	
The RBA rewriting system will be extended later and used to compare two XACML policies.

%

\section{Description of the approach}
\label{approach}

In this section, we focus on the assessment of the relationship between XACML policies. Given two policies $P_1$ and $P_2$, the process involves three steps namely:
\begin{itemize}

\item Transformation: XACML policies are mapped to boolean terms of a boolean ring while taking into account the policy and rule combining algorithms.

\item Goal: The relationship goal is transformed into a validity problem.

\item Resolution: The problem is resolved using a theorem prover based on a rewriting system.
\end{itemize}

In the sequel, we detail the three steps and we illustrate them by an example.

\subsection{From XACML to boolean expressions}

\begin{itemize}
    \item Each rule of the policy is transformed into a pair $(a,d)$ where $a$ and $d$ are boolean expressions representing the accept part and the deny part of a rule or a policy.

    \item The \textbf{Condition} of a rule is transformed to a boolean expression.
    \item If the \textbf{Condition} is missing, it is implicitly true {\sf T}.

    \item If the \textbf{Effect} of a rule is \textit{accept} and its condition is $c$, then it is transformed to $(c,{\sf F})$ (where {\sf F} represents false or an empty set).
    \item If the \textbf{Effect} of a rule is \textit{deny} and its condition is $c$, then it is transformed to $({\sf F},c)$.
    \item The condition $t$ of the \textbf{Target} is conjectured to the \textit{accept} and the \textit{deny} parts: $(t\wedge a, t\wedge d)$
\end{itemize}

The algorithm (First-Applicable, Permit Overrides, Deny Overrides, etc.), used to compose rules, produces a new one as defined \cite{ref16}. Here are some examples, where $P_1=(a_1,d_1)$ and $P_2=(a_2,d_2)$ are two policies and $a_i$ and $d_i$, $i\in\{1,2\},$ are the accept and the deny conditions of each of them:
 \begin{itemize}
    \item \textbf{First-Applicable} (denoted by $FA$): $$\begin{array}{lll}
    FA(P_1,P_2)&=&FA((a_1,d_1),(a_2,d_2))\\
    &=&(a_1\vee (a_2 \wedge \neg d_1), d_1\vee (d_2\wedge \neg a_1))
    \end{array}$$
    \item \textbf{Permit Overrides} (denoted by $PO$): $$\begin{array}{lll}
    PO(P_1,P_2)&=&PO((a_1,d_1),(a_2,d_2))\\
    &=&(a_1\vee a_2, (d_1\wedge \neg a_2)\vee (d_2 \wedge \neg a_1))
    \end{array}$$
    \item \textbf{Deny Overrides} (denoted by $PO$): $$\begin{array}{lll}
    PO(P_1,P_2)&=&PO((a_1,d_1),(a_2,d_2))\\
    &=&((a_1\wedge \neg d_2) \vee (a_2\wedge \neg d_1), d_1\vee d_2)
    \end{array}$$
\end{itemize}


\begin{example}\label{Ex1}
Consider the following XACML code that defines a policy called \texttt{SimplePolicy1} that is composed of two rules named \texttt{SimpleRule1} and \texttt{SimpleRule2} as depicted in Figure \ref{simplepolicy1}.

%

\begin{figure}%
    \centering
    \subfloat{{\includegraphics[scale=0.5]{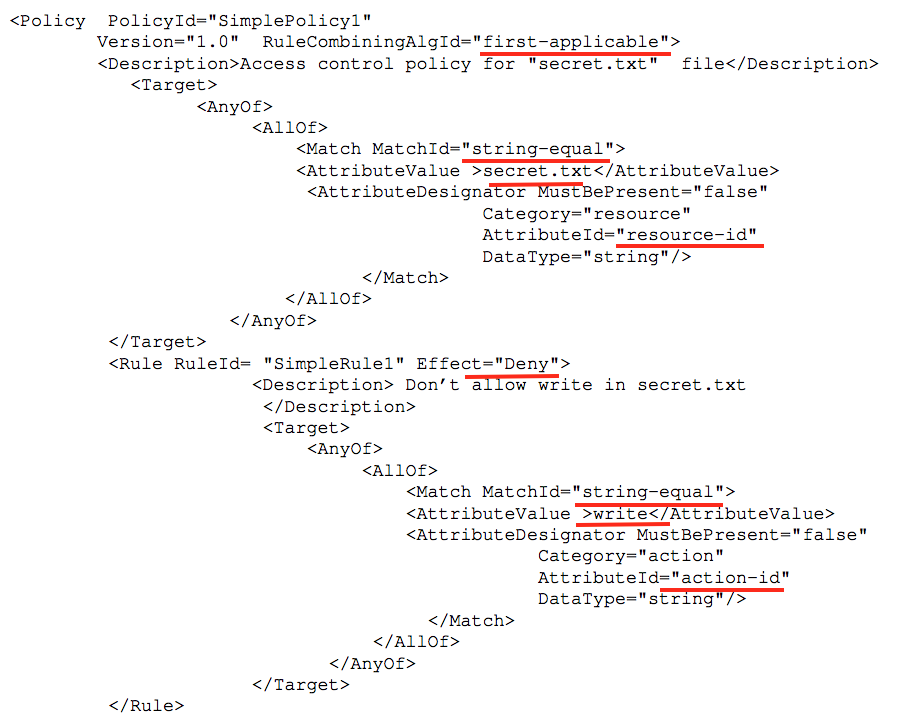} }}%
    \qquad
    \subfloat{{\includegraphics[scale=0.32]{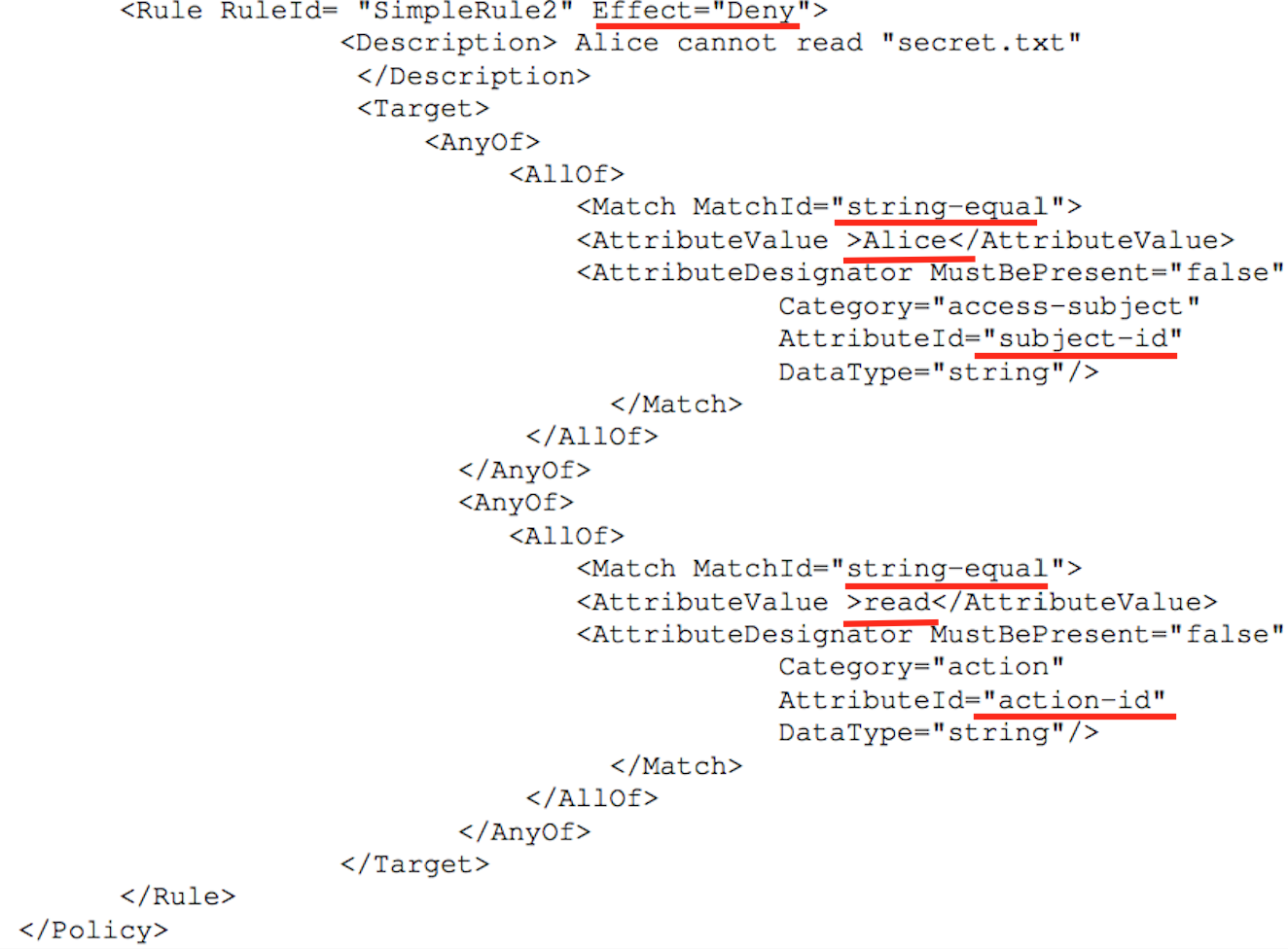} }}%
    \caption{Simple Policy 1}%
    \label{simplepolicy1}%
\end{figure}

Let $x_1$, $x_2$, $x_3$ and $x_4$ be the following variables:

 \begin{tabular}{lll}
    $x_1$&=&\texttt{string-equal(source-id, secret.txt)}   \\
    $x_2$&=&\texttt{string-equal(actionId,write)} \\
    $x_3$&=&\texttt{string-equal(subjectId,Alice)}\\
    $x_4$&=&\texttt{string-equal(actionId,read)}\\
\end{tabular}

Therefore:\\

    \texttt{SimpleRule1 }=$({\sf F},x_2)$ since \textbf{Condition}	={\sf T}, \textbf{Effect} =\texttt{Deny} and \textbf{Target}= $x_2$. \\
    \texttt{SimpleRule2 }:=$(x_3 \wedge x_4, {\sf F})$ since \textbf{Condition}	={\sf T}, \textbf{Effect} =\texttt{Permit} and \textbf{Target}= $x_3 \wedge x_4$. \\

    Since the two rules are composed using the First-Applicable algorithm, it follows that:

    \texttt{First-Applicable(SimpleRule1,SimpleRule2)}= $({\sf F}\vee(x_3 \wedge x_4\wedge \neg x_2),x_2\vee  ({\sf F}\wedge \neg {\sf F}))$\\

    Finally, since the \textbf{Target} of \texttt{SimplePolicy1} is $x_1$, the extended boolean expression representing \texttt{SimplePolicy1}, denoted by$ \lceil\texttt{SimplePolicy1}\rceil$, is:\\
     $\lceil\texttt{SimplePolicy1}\rceil=(x_1\wedge({\sf F}\vee(x_3 \wedge x_4\wedge \neg x_2)),x_1\wedge(x_2\vee  ({\sf F}\wedge \neg {\sf F})))$
 \end{example}

\subsection{From relationship analysis to validity problem}

Our policies relationship analysis is transformed into a validity test (a tautology test) in a boolean ring or a boolean algebra. Since we are interested into comparing security policies, we can easily transform different kinds of queries to tautology tests as shown in Table \ref{ComparePolicies}.
\begin{table}[!h]
    \caption{Comparing Security Policies\label{ComparePolicies}}
    \begin{center}
        \begin{tabular}{||c|c|c||}
            \hline
            \hline
            \rule[-1ex]{0pt}{2.5ex} Case & Figure & Term to prove \\
            \hline
            \hline
            &&\\
            \rule[-1ex]{0pt}{2.5ex} $P_1$ converges $P_2$ &  \includegraphics[scale=0.6]{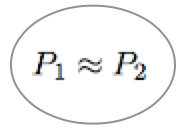}& $P_1\approx P_2$ \\
            &&\\
            \hline
            &&\\
            \rule[-1ex]{0pt}{2.5ex} $P_1$ extends $P_2$ &  \includegraphics[scale=0.6]{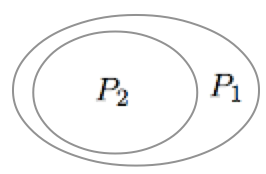}& $(P_1\Rightarrow P_2) \wedge \neg( P_2\Rightarrow P_2)$ \\
            &&\\
            \hline
            &&\\
            \rule[-1ex]{0pt}{2.5ex} $P_1$ restricts $P_2$ &  \includegraphics[scale=0.6]{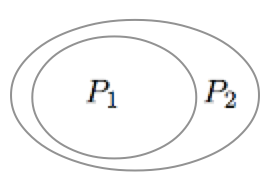}& $(P_2\Rightarrow P_1) \wedge \neg(P_1 \Rightarrow  P_2)$ \\
            &&\\
            \hline
            &&\\
            \rule[-1ex]{0pt}{2.5ex} $P_1$ shuffles $P_2$ &  \includegraphics[scale=0.6]{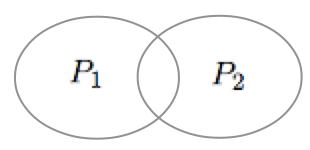}& $\neg (P_1\oplus P_2) \wedge \neg (P_2\Rightarrow P_1) \wedge \neg(P_1 \Rightarrow P_2)$ \\
            &&\\
            \hline
            &&\\
            \rule[-1ex]{0pt}{2.5ex} $P_1$ diverges $P_2$ &  \includegraphics[scale=0.6]{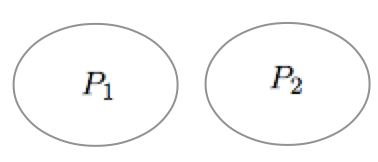}& $P_1\oplus P_2$ \\
            &&\\
            \hline
            \hline
        \end{tabular}
    \end{center}
\end{table}

\begin{example}\label{Ex2}
Suppose that we have \texttt{SimplePolicy1} of Example \ref{Ex1} and another one \texttt{SimplePolicy2} such that
$\lceil\texttt{SimplePolicy2}\rceil=(x_1\wedge({\sf F}\vee(x_4\wedge \neg x_2)),x_1\wedge(x_2\vee  ({\sf F} \wedge \neg {\sf F})))$. Suppose also that we want to prove that \texttt{SimplePolicy1} is included in \texttt{SimplePolicy2}. This goal is formalized as follows: $$\lceil\texttt{SimplePolicy1}\rceil \Longrightarrow \lceil\texttt{SimplePolicy2}\rceil$$
\end{example}

\subsection{Resolution of the problem}
\label{SR2}
To prove any tautology of Table \ref{ComparePolicies}, we propose a theorem prover that is a confluent and terminating rewriting system built mainly from RBA (Table \ref{MainRule}) and extended by the rules of Table \ref{RuleExtension}.

\begin{table}[H]
    \caption{Rules Extension\label{RuleExtension}}
    \begin{center}
        	$\begin{array}{rlll}
        (0,0)&\rightarrow &0& T_{0} \\
        (0,1) &\rightarrow &0 & T_{1}\\
        (1,0)&\rightarrow& 0 & T_{2}\\
        (1,1)& \rightarrow &1 &  T_{3}\\
        \neg (\varphi_1,\psi_1)&\rightarrow& (\neg \varphi_1, \neg \psi_1)& T_{4} \\
        (\varphi_1,\psi_1)\approx (\varphi_2,\psi_2)&\rightarrow& (\varphi_1\approx \varphi_2,\psi_1 \approx \psi_2)& T_{5} \\
        (\varphi_1,\psi_1)\vee (\varphi_2,\psi_2)&\rightarrow& (\varphi_1\vee \varphi_2,\psi_1 \vee \psi_2)& T_{6} \\
        (\varphi_1,\psi_1)\wedge (\varphi_2,\psi_2)&\rightarrow& (\varphi_1\wedge \varphi_2,\psi_1 \wedge \psi_2)& T_{7} \\
        (\varphi_1,\psi_1)\oplus (\varphi_2,\psi_2)&\rightarrow& (\varphi_1\oplus \varphi_2,\psi_1 \oplus \psi_2)& T_{8} \\
        (\varphi_1,\psi_1)\Longrightarrow (\varphi_2,\psi_2)&\rightarrow& (\varphi_1\Longrightarrow \varphi_2,\psi_1\Longrightarrow \psi_2)& T_{9} \\
        \end{array}$
    \end{center}
\end{table}

The previous rewriting rules should be applied under commutativity and associativity of $\ast$ and $\oplus$, i.e:

\begin{center}
    $\begin{array}{llll}
    Commutativity &x \oplus y&= & y \oplus x \\
    &x \ast y &=& y \ast x\\
    Associativity &(x \oplus y) \oplus z&=&x \oplus (y \oplus z)\\
    &(x \ast y) \ast z&=&x \ast (y \ast z)\\
    \end{array}$
\end{center}

The rule $T_0,\ldots,T_3$ states that a pair $(\phi,\psi)$ is considered as a tautology only if both $\phi$ and $\psi$ are tautologies. Suppose for example that we want to prove that $(\varphi_1\Longrightarrow \varphi_2,\psi_1\Longrightarrow \psi_2)$, this will be true only when each part of the pair is evaluated to true.

\begin{example}\label{Ex3}
    We apply here the rewriting system to prove the goal of Example \ref{Ex2}. Since $\lceil\texttt{SimplePolicy1}\rceil$ has the form of $(\phi_1,\psi_1)$ and $\lceil\texttt{SimplePolicy1}\rceil$ has the form of $(\phi_2,\psi_2)$, then according to rule $T_9$, proving that $\lceil\texttt{SimplePolicy1}\rceil \Longrightarrow \lceil\texttt{SimplePolicy2}\rceil$ implies proving that $(\phi_1\Longrightarrow\phi_2,\psi_1\Longrightarrow\psi_2)$. Hereafter, we prove that $\phi_1\Longrightarrow\phi_2$. A similar proof can be done for $\psi_1\Longrightarrow\psi_2$.

\begin{align*}
&\phi_1\Longrightarrow\phi_2\\
 \rightarrow& \{  \mbox{From the definition $\phi_1$ and $\phi_2$ } \}\\
    &x_1\wedge({\sf F}\vee((x_3 \wedge x_4) \wedge \neg x_2)) \Longrightarrow x_1\wedge({\sf F}\vee(x_4 \wedge \neg x_2))\\
    \rightarrow& \{ R_0 \mbox{(Two times) } \}\\
    &x_1\wedge(0\vee((x_3 \wedge x_4) \wedge \neg x_2)) \Longrightarrow x_1\wedge(0\vee(x_4 \wedge \neg x_2))\\
    \rightarrow& \{ R_2 \mbox{(Two times) } \}\\
    &x_1\wedge(0*((x_3 \wedge x_4) \wedge \neg x_2)\oplus 0 \oplus (x_3 \wedge x_4 \wedge \neg x_2) ) \Longrightarrow\\ &x_1\wedge(0*(x_4 \wedge \neg x_2)\oplus 0 \oplus (x_4 \wedge \neg x_2))\\
    \rightarrow& \{ R_{12} \mbox{(Two times)  } + \mbox{Commutativity of * }\}\\
    &x_1\wedge(0\oplus 0 \oplus ((x_3 \wedge x_4)-x_2) ) \Longrightarrow x_1\wedge(0\oplus 0 \oplus (x_4-x_2))\\
    \rightarrow& \{ R_{7} \mbox{(Many times)  } + \mbox{Commutativity of $\oplus$ }\}\\
    &x_1\wedge ((x_3 \wedge x_4)\wedge \neg x_2)  \Longrightarrow x_1\wedge (x_4 \wedge \neg x_2)\\
    \rightarrow& \{ R_{3} \mbox{(Many times)  } \}\\
    &x_1* ((x_3 * x_4) * \neg x_2)  \Longrightarrow x_1* (x_4 * \neg x_2)\\
    \rightarrow& \{ R_{6} \mbox{(Many times)  } \}\\
    &x_1* ((x_3 * x_4)* ( x_2 \oplus 1))  \Longrightarrow x_1* ( x_4 * (x_2\oplus 1))\\
    \rightarrow& \{ R_{12} \mbox{(Many times)  } \}\\
    &x_1*x_3 * x_4* x_2\oplus  x_1* x_3 * x_4  \Longrightarrow x_1* x_4 * x_2 \oplus x_1* x_4\\
    \rightarrow& \{ R_{4}  \}\\
    & (x_1*x_3 * x_4* x_2 \oplus  x_1* x_3 * x_4) *  (x_1* x_4 * x_2 \oplus x_1* x_4) \oplus \\
    & (x_1*x_3 * x_4* x_2 \oplus  x_1* x_3 * x_4) \oplus 1\\
    \rightarrow&  \{ R_{12} (\mbox{Many times  }) + \mbox{Commutativity of $\oplus$ } + R_{10} \mbox{(Many times)} \}\\
    & x_1* x_2*x_3 * x_4 \oplus  x_1* x_3 * x_4 \oplus            \\
    & (x_1* x_2*x_3 * x_4 \oplus  x_1* x_3 * x_4) \oplus 1\\
    \rightarrow&  \{ R_{7}+ R_{8}  +\mbox{Commutativity de $\oplus$ }  \}\\
    &1 \\
    \end{align*}

\end{example}

\subsection{Termination of the Theorem Prover }
\label{termination}

In this section, we prove the termination of the proposed rewriting system using AProVE \cite{Aprove2004}; an online system for automated termination and complexity proofs of terms rewriting systems. This tool proves the termination of our rewriting system and gives the proof of this termination. AProVE takes as input a set of rewriting rules and a set of equations and returns as output an ordering relation according to which the system is proved to be terminating when this is the case. Otherwise, it returns ``fail".

\begin{figure}[H]
    \centering
    \includegraphics[scale=0.3]{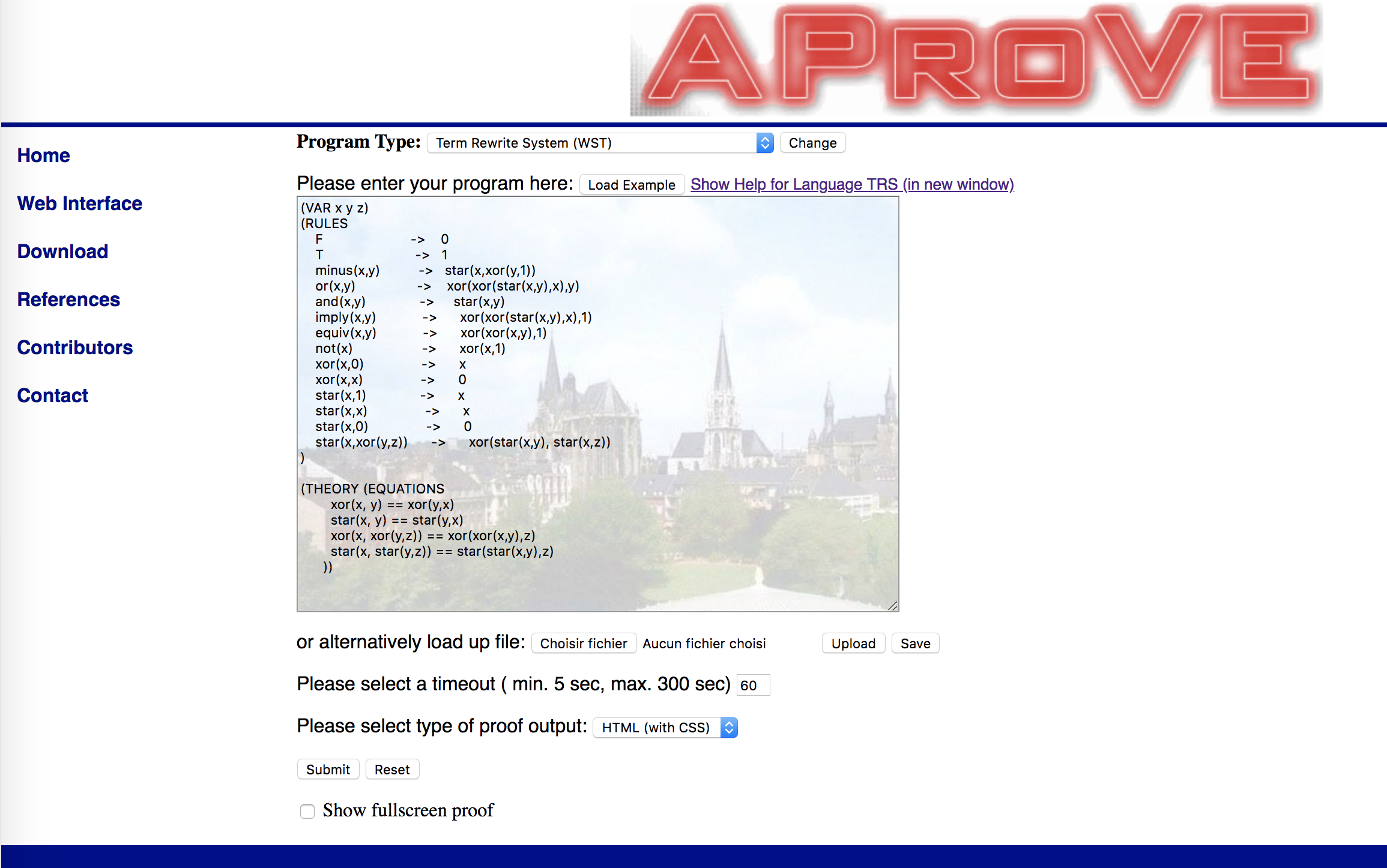}\\
    \caption{AProVE Input}
    \label{aprove1}
\end{figure}

As shown in Figure \ref{aprove1}, we changed the names of functions to comply with the syntax of the tool:

$$
\begin{array}{|c|c|}
\hline
\hline
\mbox{Rewriting System  Operator} & \mbox{AProVE Operator}\\
\hline
\hline
\vee & \texttt{or}\\
\wedge & \texttt{and}\\
\oplus & \texttt{xor}\\
\star & \texttt{star}\\
\neg & \texttt{not}\\
\Longrightarrow & \texttt{imply}\\
\approx & \texttt{equiv}\\
\hline
\end{array}$$

\begin{figure}[h!]
    \centering
    \includegraphics[scale=0.35]{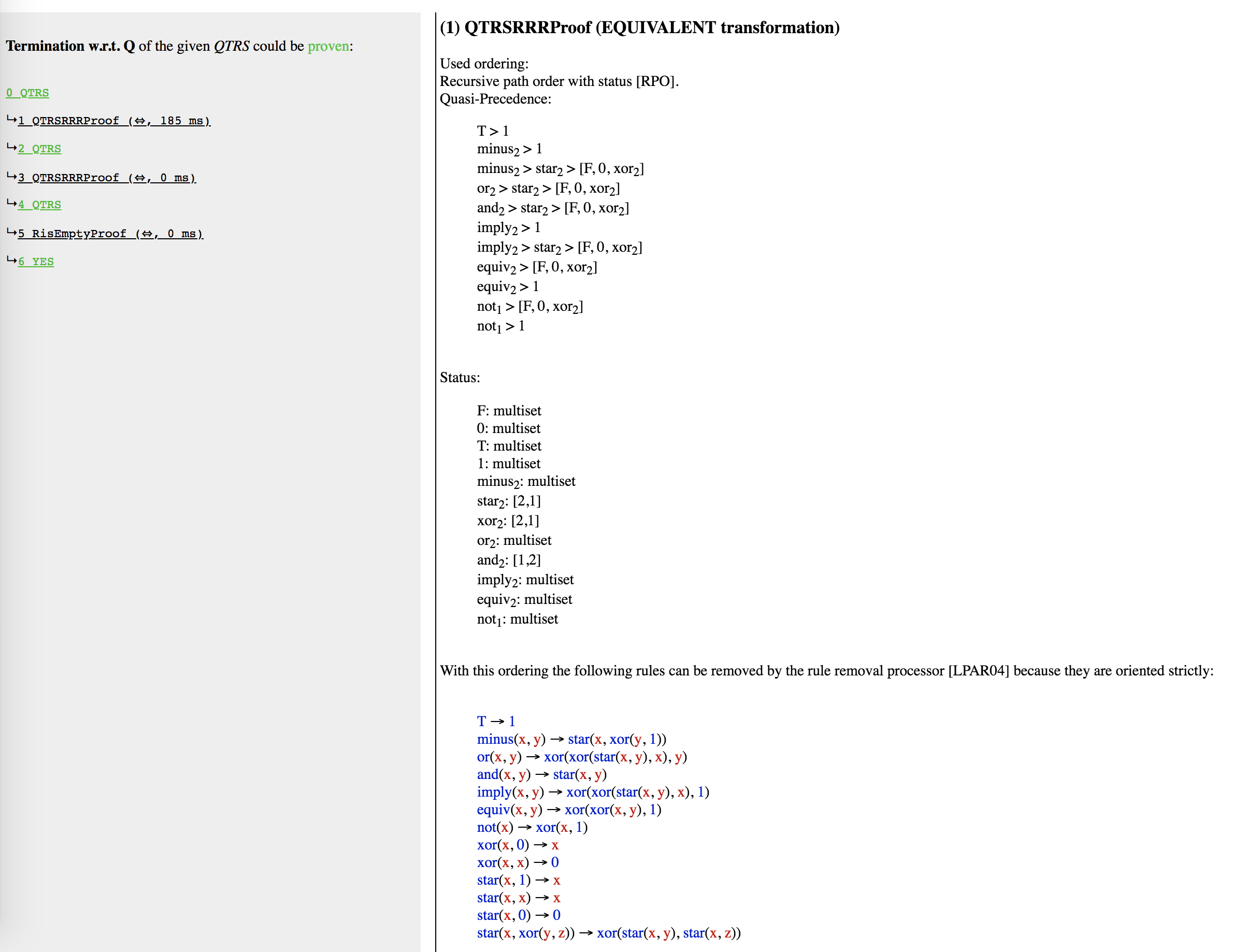}\\
    \caption{AProVE Output}
    \label{aprove2}
\end{figure}

As outlined by Figure \ref{aprove2}, using a Recursive Path Order \cite{Dershowitz82} (RPO) with status, the tool proves that our rewriting system is terminating. The tool shows that, based on the following partial ordering, the left size of any rule is greater than its right size with respect to RPO.

\begin{verbatim}
F > [0, xor]
T > [1, not] > [0, xor]
minus > [1, not] > [0, xor]
minus > star > [0, xor]
or > star2> [0, xor]
and > star > [0, xor]
imply > [1, not] > [0, xor]
imply > star > [0, xor]
equiv > [1, not] > [0, xor]
\end{verbatim}

\section{Prototype and Experimental Evaluation}
\label{exp}

\begin{figure}[h!]
  \centering
  \includegraphics{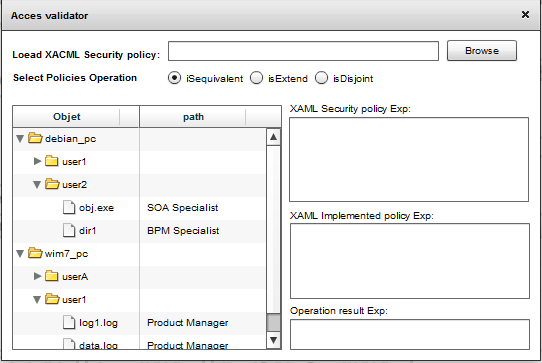}\\
  \caption{Policy relationship interface}
  \label{polcompinterface}
\end{figure}

To analyze the performance of our policies relationship assessment approach, we developed a prototype using Java and XACML. More precisely, we used Tom \cite{DBLP:conf/rta/BallandBKMR07}; a pre-compiler that enables us to implement our rewriting system and transformation rules. Figure \ref{polcompinterface} shows the interface of policies relationship assessment that we have implemented as part of the whole tool.

%
%

\subsection{Performance analysis}

Our approach transforms the problem of comparing two XACML policies to an extended version of SMT (Satisfiability Modulo Theories) problem.  After that we can use any SMT technique to solve the problem. We preferred the use of an SMT based on boolean rings since it has been shown in  \cite{Dershowitz04booleanring} to be efficient in many cases and it even gives polynomial-time results for restricted classes of boolean ring formulae.

The satisfiability testing using boolean rings is NP-hard \cite{BHR84}.  The major problem comes from the distributivity law ($x\star (y \oplus z)=x\star y \oplus x\star z$) that can cause the length of a boolean term to grow exponentially in the worst case. Accordingly, many simplifications at intermediate stages are available in \cite{Dershowitz04booleanring,MarquesSilva2000} for the sake of reducing the likelihood of suffering from the exponential aspect.

To evaluate our approach, we developed a tool that generates XACML policies with a random number of rules. The experiments presented in this section are performed 10 times on a machine Core i7, having 2.20 Ghz as processor speed and 8 GB of RAM with a macOS Mojave. The average of the obtained results is then computed.

We start by doing some experiments to evaluate the first step of our approach, which is the translation of  XACML policies into a boolean expression. Figure \ref{time} shows the performance of the system for each type of relationship. The trends of the five curves are almost the same. Furthermore, the curves are almost linear with respect to the number of rules in a policy.

\begin{figure}[h!]
  \centering
 \begin{tikzpicture}
\begin{axis}[
height=9cm,
width=9cm,
grid=major, legend pos=outer north east,
ylabel=Milliseconds,xlabel=Num rules,cycle list name=exotic
]
\addplot coordinates {
(4,57)
(8,85)
(12,106)
(16,121)
(20,176)
};
\addlegendentry{Diverge}
\addplot coordinates {
(4,49)
(8,77)
(12,114)
(16,116)
(20,161)
};
\addlegendentry{Converge}
\addplot coordinates {
(4,52)
(8,84)
(12,110)
(16,121)
(20,172)
};
\addlegendentry{Extend}
\addplot coordinates {
(4,47)
(8,77)
(12,111)
(16,119)
(20,170)
};
\addlegendentry{Shuffle}
\addplot coordinates {
(4,53)
(8,77)
(12,106)
(16,117)
(20,173)
};
\addlegendentry{Restrict}
\end{axis}
\end{tikzpicture}
  \caption{CPU Execution average time for each type of relationship}
  \label{time}
\end{figure}

\subsubsection{Preprocessing time}

We evaluate here the time required to rewrite a policy as a term in a boolean ring for a pair of policies. The average number of parameters (resource, action and subject) varies between 20 and 100 for each policy and the number of rules
varies between 8 and 32. The preprocessing time is linear with respect to the number of rules and parameters as depicted
in Figure \ref{PreprocessingTime}. More precisely, if
each policy has 32 rules and 100 parameters, the preprocessing time is 90\% the global response time. Hence, the global relationship response
time is dominated by the preprocessing module.

\begin{figure}[h!]
 \begin{center}
\begin{tikzpicture}
\begin{axis}[
legend pos= north west 
,ybar
,ylabel=Preprocessing Time(s),xlabel= Number of parameters
,xtick={0,20,40,60,80,100} 
 ,xmin=10,xmax=110
,xtick align=inside
,axis x line=bottom
,axis y line=left
,x axis line style={|-|} 
,bar width=8pt 
,]
\addplot coordinates {
 (20,1.165)
(40,2.795)
(60,4.872)
(80,7.687)
(100,12.041)

};
\addplot coordinates {
 (20,2.493)
(40,5.936)
(60,11.269)
(80,15.018)
(100,23.478)
};
\addplot coordinates {
  (20,5.596)
(40,13.516)
(60,25.982)
(80,43.061)
(100,59.081)
};
\legend{8 Rules, 16 Rules, 32 Rules}
\end{axis}
\end{tikzpicture}
\end{center}
  \caption{Preprocessing time for a pair of policies}
  \label{PreprocessingTime}
\end{figure}

\subsubsection{Response time}

Now, we evaluate the total time for policies relationship assessment. Figure \ref{pairs} shows the
time variation while increasing the number of compared policies.

\begin{figure}[h!]
 \begin{center}
\begin{tikzpicture}
\begin{axis}[
legend pos= north west 
,ybar
,ylabel=Total Response Time(s),xlabel= Number of parameters
,xtick={0,20,40,60,80,100} 
,xmin=10,xmax=110 
,xtick align=inside
,axis x line=bottom
,axis y line=left
,x axis line style={|-|} 
,bar width=8pt 
,]
\addplot coordinates {
(20,1.248)
(40,2.823)
(60,4.856)
(80,7.981)
(100,10.689)
};
\addplot coordinates {
(20,2.498)
(40,6.116)
(60,10.648)
(80,17.479)
(100,27.280)
};
\addplot coordinates {
(20,6.116)
(40,13.402)
(60,29.246)
(80,40.292)
(100,65.151)
};
\legend{8 Rules, 16 Rules, 32 Rules}
\end{axis}
\end{tikzpicture}
\end{center}
  \caption{Total response time to compare a pair of policies}
  \label{pairs}
\end{figure}

Knowing that the number of pairs is generally between 20 and 100 \cite{ref10}, our approach has a reasonable response time.
Another experiment is done by fixing the number of elements in each policy and varying the number of policy pairs to
be compared. The policies have 10 parameters in average. We did the experiment for 8 to 16 rules as shown in Figure \ref{rules}.

\begin{figure}[h!]
  \centering
 \begin{tikzpicture}
\begin{axis}[
legend pos=north west 
,height=9cm,
width=9cm,
grid=major,
ylabel=Seconds,xlabel=Num pairs]
\addplot coordinates {
(4,1.044)
(8,2.099)
(16,4.308)
(32,8.450)
(64,16.776)
(128,31.959)
};
\addlegendentry{8 Rule}
\addplot coordinates {
(4,2.104)
(8,4.468)
(16,8.600)
(32,17.567)
(64,35.14)
(128,70.99)
};
\addlegendentry{16 Rule}
\end{axis}
\end{tikzpicture}
  \caption{CPU execution average time}
  \label{rules}
\end{figure}

We observe that the time required to compare few hundreds of policies is a few seconds. The minimal response time is 1 s and
the maximal is 1 mn. This shows the feasibility and scalability of our approach.

\subsection{Comparison with similar SMT approaches}

In this section, we evaluate the performance (time and memory) of our approach versus the best so far SMT based approach \cite{DBLP:journals/compsec/TurkmenHRZ17} on real and synthetic policies.

\subsubsection{Real Policies}
		
Figure \ref{Real Time Figure} displays the total time (translation and solving) of both approaches on three real world policies: \textbf{GradeMan}\footnote{\url{http://www.margrave-tool.org/v1+v2/margrave/versions/01-01/examples/college/}}, \textbf{KMarket}\footnote{\url{https://svn.wso2.org/repos/wso2/trunk/commons/balana/modules/balana-samples/kmarket-trading-sample/resources/}} and \textbf{Continue}\footnote{\url{http://websitehive.com/wsphp/distance/upload/continue-a.xml}}. Our approach achieved much better results on \textbf{GradeMan} and \textbf{KMarket} policies, and close execution time for the \textbf{Continue} policy compared to \cite{DBLP:journals/compsec/TurkmenHRZ17}. The total time for our approach ranges from 18 ms up to 3925 ms. The results vary depending on the number of policy sets, policies, rules, and conditions. For \cite{DBLP:journals/compsec/TurkmenHRZ17}, the total time ranges from 249 ms to 3715 ms. The displayed table provides more details about the comparison, where the number of policy sets, policies and rules are shown for each policy.

\begin{small}
\begin{figure}[h]
		
    \begin{tabular}{cc}
 		
			\centering
			\includegraphics[width=\textwidth]{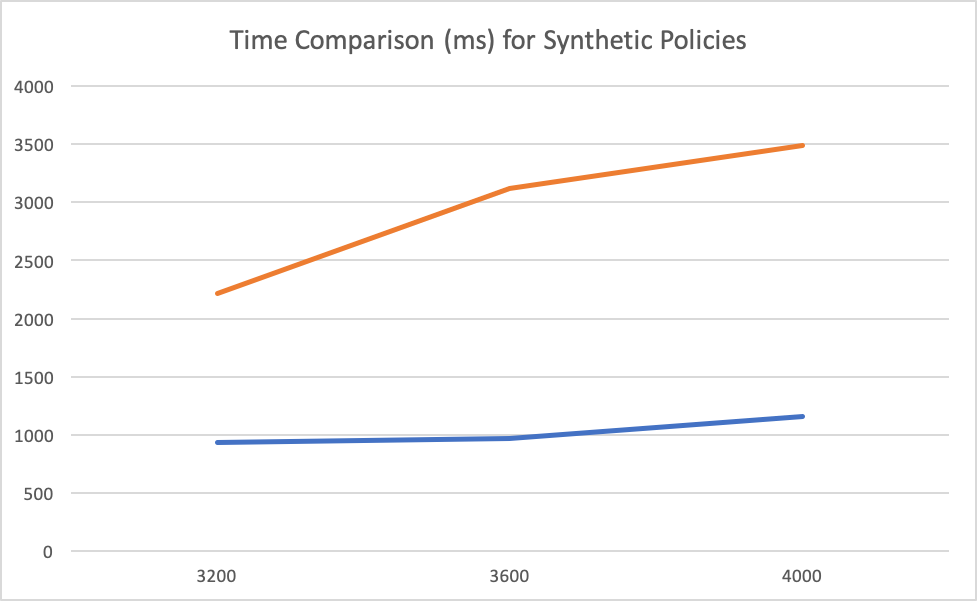}
		
    &\\
			\begin{tabular}{|c|c|c|c|c|c|}
				\hline
				Policy    & \#PS & \#P & \#R & Our approach (ms) & SMT based approach \cite{DBLP:journals/compsec/TurkmenHRZ17} (ms) \\ \hline
				GradeMan & 11   & 5   & 5   & 102            & 326               \\ \hline
				KMarket   & 1    & 3   & 12  & 18             & 249               \\ \hline
				Continue  & 111  & 266 & 298 & 3925           & 3715              \\ \hline
			\end{tabular} &\\

\end{tabular}

\caption{Time comparison of real policies}
\label{Real Time Figure}

\end{figure}
\end{small}
		
\begin{figure}[h]
		
    \begin{tabular}{cc}
 		
			\centering
			\includegraphics[width=\textwidth]{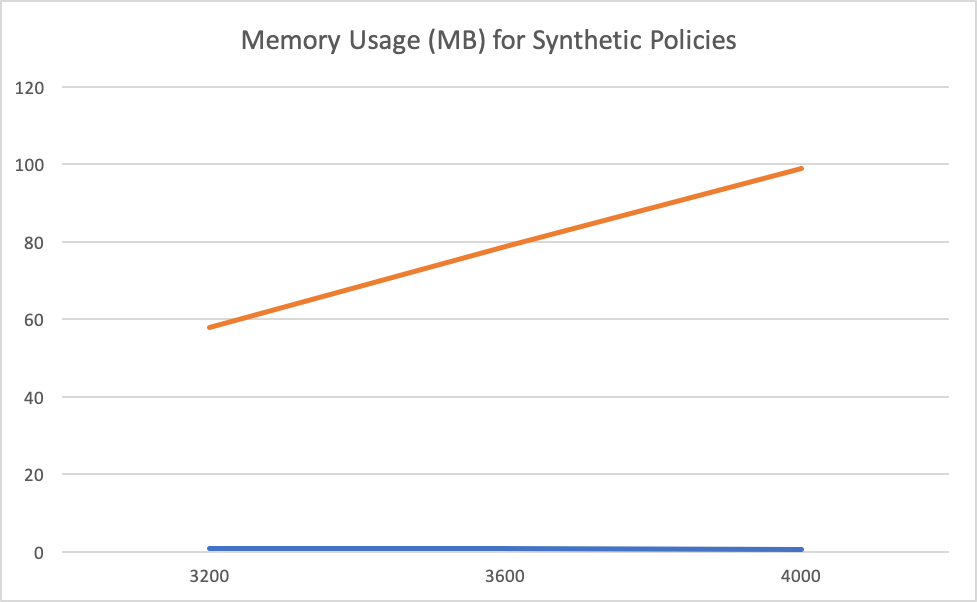}
		
    &\\
			\begin{tabular}{|c|c|c|c|c|c|}
				\hline
				Policy    & \#PS & \#P & \#R & Our approach (MB) & SMT based approach \cite{DBLP:journals/compsec/TurkmenHRZ17} (MB) \\ \hline
				GradeMan & 11   & 5   & 5   & 0.56             & 6.5                 \\ \hline
				KMarket   & 1    & 3   & 12  & 0.56             & 5.2                 \\ \hline
				Continue  & 111  & 266 & 298 & 1.29             & 223.8               \\ \hline
			\end{tabular} &\\

\end{tabular}

\caption{Memory comparison on real world policies}
\label{Real Memory Figure}

\end{figure}

Figure \ref{Real Memory Figure} outlines the memory usage (in MB) of both approaches on the three aforementioned real world policies. Our approach achieves better results for all these policies. In fact, the memory usage ranges from 0.56 MB to 1.3 MB. For \cite{DBLP:journals/compsec/TurkmenHRZ17}, the memory consumption ranges from 5.2 MB to 223.8 MB.

\subsubsection{Synthetic Policies}

In this section, we evaluate the performance (time and memory) of both approaches on synthetic policies. Performance assessment was conducted on three synthetic policies P1, P2, and P3 \cite{mourad2015sba} having respectively \textbf{3200}, \textbf{3600}, and \textbf{4000} rules. The policies are created in such a way that they are exhaustive, i.e., all rules are set to deny with a policy combining algorithm ``permit-overrides".

\begin{figure}[h]
		
    \begin{tabular}{cc}
 		
			\centering
			\includegraphics[width=\textwidth]{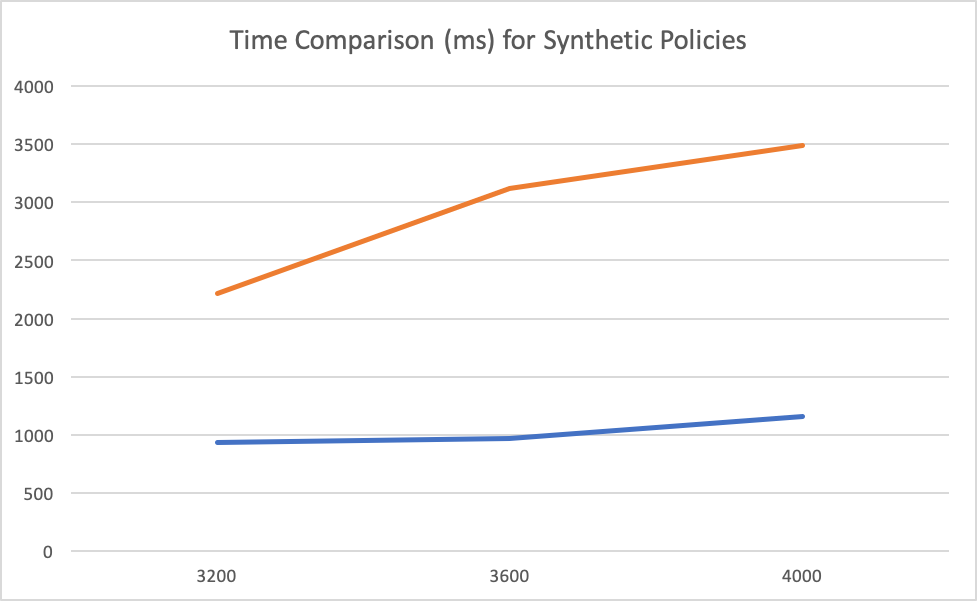}
		
    &\\
			\begin{tabular}{|c|c|c|c|c|c|}
				\hline
				Policy    & \#PS & \#P & \#R & Our approach (ms) & SMT based approach \cite{DBLP:journals/compsec/TurkmenHRZ17} (ms) \\ \hline
				P1  & 1    & 80   & 3200   & 932            & 2213               \\ \hline
				P2   & 1    & 90   & 3600   & 973           & 3115               \\ \hline
				P3  & 1   & 100   & 4000  &  1156           & 3486              \\ \hline
			\end{tabular} &\\

\end{tabular}

\caption{Time comparison on synthetic policies}
\label{Synthetic Time Figure}
\end{figure}

Figure \ref{Synthetic Time Figure} presents the total time (translation/solving) of both approaches on synthetic policies. The total time for our approach ranges from 932 ms to 1156 ms while for \cite{DBLP:journals/compsec/TurkmenHRZ17} the total time ranges from 2213 ms to 3486 ms.

\begin{figure}[h]
		
    \begin{tabular}{cc}
 		
			\centering
			\includegraphics[width=\textwidth]{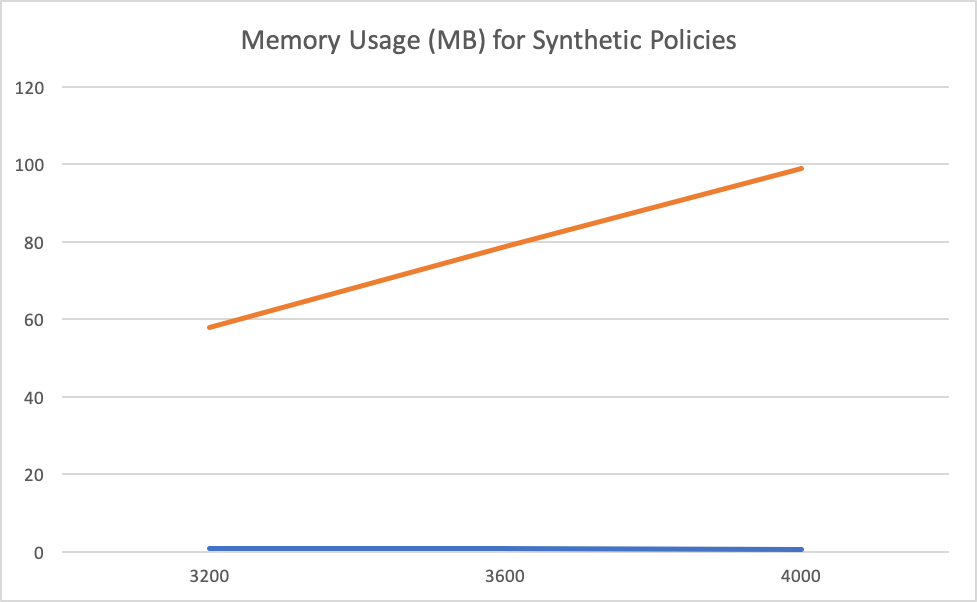}
		
    &\\
			\begin{tabular}{|c|c|c|c|c|c|}
				\hline
				Policy & \#PS & \#P & \#R & Our approach (MB) & SMT based approach \cite{DBLP:journals/compsec/TurkmenHRZ17} (MB) \\ \hline
				P1   & 1   & 80   & 3200    & 0.891            & 57.89              \\ \hline
				P2   & 1  & 90    & 3600    & 0.81             & 78.73              \\ \hline
				P3   & 1  & 100   & 4000    & 0.65             & 98.98              \\ \hline
			\end{tabular} &\\

\end{tabular}

\caption{Memory comparison on synthetic policies}
\label{Synthetic Memory Figure}
\end{figure}

Figure \ref{Synthetic Memory Figure} outlines the memory usage for both approaches on synthetic policies. For our approach, the memory usage ranges from 0.65 MB up to 0.891 MB while for \cite{DBLP:journals/compsec/TurkmenHRZ17} the memory usage ranges from 57.89 up to 98.98 MB. The displayed table includes a more detailed comparison.

The results confirm that rewriting policies based on boolean rings enjoys better
performance and memory cost than SMT-based approaches.

\section{Related Work}
\label{relwork}

Policies analysis was lately subject to intensive research works. We review in what follows these
works and we mainly focus on those related to the assessment of policies relationship.

\begin{figure}
\begin{center}
\scalebox{0.40}{
\includegraphics{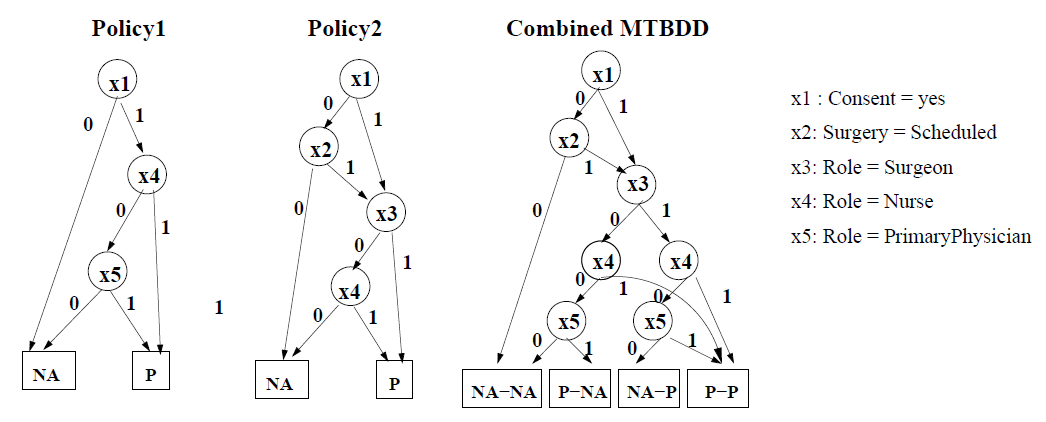}}
\end{center}
\caption{Policy MTBDD}
 \label{PolMTBDD}
\end{figure}

Exam \cite{ref10} is a complete environment for the analysis and management of access control policies. It
supports the acquisition, search, analysis of relationship, and integration of security policies. This method uses the Multiple Terminal
Binary Decision Diagram (MTBDD) as a representation of a policy. The MTBDD of a security policy is an acyclic directed graph whose internal nodes
represent boolean predicates that correspond to: $S$ (subject), $A$ (action), $R$ (resource), $C$ (condition), and the components
of a policy in which the terminals can be one of \{permit, deny, not applicable\} representing the effect of the policy on requests. In such diagram, the paths represent the rules of a policy. The MTBDDs of different policies are then combined to derive a single MTBDD in which each terminal corresponds to a n-tuple $<e_1,e_2,\ldots,e_n>$ that represents the effect of the policy on all requests. By browsing the paths leading to the terminals of the combined MTBDD, users can extract the set of the requests, which have common or different authorizations in a given set of policies. Also, users can deduce all the requests authorized by the policies by browsing all the paths that lead to the terminal $<Permit, Permit>$ in the combined MTBDD. For example, imagine a scenario where a patient wishes to transfer his medical records from hospital $X$ to hospital $Y$ each of which has its own policy, $P_1$ and $P_2$ respectively. The policy $P_1$ allows access to medical records if the patient has given his consent and if the subject who needs to access is either a nurse or a doctor. The policy $P_2$ allows access to medical records if the patient has given his consent or a surgical intervention is planned and if the subject is either a surgeon or a nurse. Before transferring the medical records, the patient must ensure that the security policy of hospital $Y$ offers the same level of security as that of hospital $X$. Two security policies are considered to be the same if their combining MTBDD has leaves labeled with either NA-NA or P-P. Leaves of the form NA-P, P-NA show divergences between the policies. An analysis of relationship between $P_1$ and $P_2$ can be used to ensure such a requirement. Figure \ref{PolMTBDD} shows the MTBDDs corresponding to $P_1$, $P_2$ and their combination. Indeed, the paths to the terminal $P - P$ in the combined MTBDD pinpoint the relations between the two policies, whereas the remaining paths show some differences. The patient may use this information before deciding to transfer his medical records. Although this approach is precise, it could not be directly applied on a set of XACML security policies (PolicySet) because their global semantics depends on their combining algorithm (Permit-overrides, Deny-overrides, etc.).  To apply the approach, the set of the security policies and their combining algorithms should be first transformed to a simple boolean expression.\\


Margrave \cite{ref12} contains, in addition to the verification component of properties, a system for the analysis of impact change. It considers two policies and summarizes the differences between them. Margrave allows users to represent the security policies in the form of trees using the same procedure as the tree that is manipulated by EXAM. The tree resulting from this combination can be further analyzed and verified, with the tool Margrave as described in \cite{ref12}. Margrave defines a diagram of decisions called a decision diagram of analysis change or a Common Multiple Terminal Binary Decision Diagram (CMTBDD) whose representation shows the combinations of changes occurring between the policies as well as the differences between them. This tool has the disadvantage of not being able to model the structures of different specification languages of security policies and is limited to Ponder and XACML languages in addition to the difficulty to obtain analysis results for complex specifications. Our algebraic setting can capture different Attribute Based Access Control (ABAC) specification languages including XACML. Furthermore, the aforementioned works rely on BDD and MTBDD to determine the policies relationship. However, the main issue with mapping policies to BDD or MTBDD is the lack of scalability as pinpointed in \cite{FACPL-2015}.

In \cite{ref15}, Backes et al. proposed an algorithm to verify the refinement of business confidentiality policies. The proposed policy refinement checks if a policy is a subset of the other. However, their study is based on EPAL rather than XACML. This is an important difference with our work because the sole rule combining algorithm that is considered in their work is the First-one-applicable.

Jebbaoui et al. \cite{Jebbaoui2015} devised a new set-based language called SBA-XACML to capture the complex structures in XACML and endowed it with semantics to detect access flaws, conflicts and redundancies between policies. Their work covers some rules combining algorithms (e.g., Permit overrides and First Applicable). The correctness of the approach is not discussed in that paper.

In \cite{DBLP:conf/post/TurkmenHRZ15,DBLP:journals/compsec/TurkmenHRZ17} Turkmen et al. proposed to use Satisfiability Modulo Theories (SMT) to analyze XACML policies. They formalized XACML using predicates and they encoded them in SMT \cite{SMT-BDDT2009}. They also devised a query language for a variety of policy checking and they implemented their approach in a prototype called X2S \cite{DBLP:conf/ccs/TurkmenHZ14}.

In \cite{DBLP:journals/corr/MargheriPT15}, Margheri et al. proposed an automated verification of XACML policies based on a constraint solver software. The security policies are first formalized using the FACPL language \cite{FACPL-2015}. Then, a semantic-based approach is followed to capture FACPL policies. Furthermore, they formalized various properties on the structure of policies to characterize the relationships with the behaviors that different policies enforce. Finally, they use constraint solver tools for the automated verification of security and structural properties.

Our work proposes an algebraic approach to formalize XACML policies and assess the relationship between policies. More precisely, we provide a formal framework that allows the users to reason formally about XACML policies and prove the type of relationship between policies. Compared to the approaches that use a formal language to capture XACML policies such as \cite{DBLP:conf/post/TurkmenHRZ15,DBLP:journals/compsec/TurkmenHRZ17} and \cite{DBLP:journals/corr/MargheriPT15}, our approach relies on a simple and compact algebraic language whose foundations come from boolean rings rather than a complex language with a tailored semantics. Furthermore, our approach spans over a rewriting system to assess the policies relationship. Using a rewriting system gives us an framework that can address more problems than what SMT and BDD can do. Also, as shown in \cite{Dershowitz04booleanring}, an SMT based on a boolean ring and rewriting can be efficient in many cases and it even gives polynomial-time results for restricted classes of boolean ring formulae.  Moreover,  whether we use BDD, SMT or reciting techniques, we need a preliminary step that transforms an XACML policy to a boolean formula. This step should handle most of XACML rules and its combining algorithms, which is not the case of many existing approaches.


\section{Conclusion and future work}
\label{conc}

We presented a new approach for the assessment of policies relationship that is based on a rewriting system. Each policy is mapped into a term in a boolean ring based on pre-defined transformation rules. We presented an example to show how to compare two XACML policies based on our approach. We proved the termination of the rewriting system based on a tool called AProVE. We implemented the approach and we provided its experimental performance analysis. Our future work will be oriented towards the optimization of the proposed approach and its extension to deal with more rule combining algorithms.

\section*{References}


\end{sloppypar}
\end{document}